\documentclass[showpacs,twocolumn,amsmath,amssymb]{revtex4}
\usepackage{graphicx}
\usepackage{setspace}
\usepackage{epsfig}
\begin{document}
\title{Extinction of Populations and the Schr\"{o}dinger Equation: \\Analytic Calculations for Abrupt Transitions}
\author{Niraj Kumar and V. M. Kenkre} 
\address{Consortium of the Americas for Interdisciplinary Science and
         Department of Physics and Astronomy, University of New Mexico,
         Albuquerque, NM 87131, USA}
\begin{abstract}
We study bifurcations in a spatially extended nonlinear system representing population dynamics with the help of analytic calculations based on the time-independent Schr\"{o}dinger equation for a quantum particle subjected to a uniform gravitational field. Despite the linear character of the Schr\"{o}dinger equation, the result we obtain helps in the understanding of the onset of abrupt transitions leading to extinction of biological populations. The result is expressed in terms of Airy functions and sheds light on the behavior of bacteria in a Petri dish as well as of large animals such as rodents moving over a landscape. \\
\end{abstract}
\maketitle
%{PACS number: 87.23.Cc}
\section{Introduction}
The purpose of this paper is two-fold: to show how standard Schr\"{o}dinger equation analysis familiar in a simple physics situation may be used to understand an important problem in biology, and to study, via such an analysis, the effect of spatial variation of resources on the ecologically relevant problem of extinction of populations of biological species. Extinction in biology is intimately related, as we shall show below, to abrupt transitions in physical systems, and is important whether the focus is on the dynamics of bacteria in a Petri dish \cite{Lin,Perry,Dahmen,Shnerb,KK,
KN} or of the movement of larger animals such as rodents involved in the spread of epidemics 
\cite{Terry,AK,Camelo,NK}. 

Consider the logistic equation
\begin{equation}
\frac{du(t)}{dt}=au(t)-bu^2(t)
\label{logistic}
\end{equation}
which may describe the time dependence of the density $u$ of a spatially uniform population subject to a growth rate $a$ and a resource competition term controlled by the parameter $b$. The steady state density undergoes a transcritical bifurcation at $a=0,$ as $a$ is varied from negative to positive values. The physical meaning of the bifurcation is straightforward. Whereas for positive $a$ the stable steady state value of $u$ is $a/b$, for negative $a$ the  density vanishes in the steady state because  the growth rate is then really a death rate. Introduction of space as a mere parameter $X$ brings in no new behavior. 

A new aspect does enter, however, if spatial \emph{derivatives}  are introduced into Eq. (\ref{logistic}) through, for instance, diffusion terms which represent the motion of the members of the species from one spatial location to another.  It is this situation, represented by a Fisher equation \cite{Murray, Kot}with spatially varying growth rate $a(X)$ and  diffusion constant $D$,
\begin{equation}
\frac{\partial u(X,t)}{\partial t}=a(X)u(X,t)-bu^2(X,t)+D\frac{\partial ^2u(X,t)}{\partial X^2}
\label{fisher}
\end{equation}
that is the subject of the present paper. We study the 1-d  case for simplicity. As expected, the steady state $u(X)$ has a tendency to follow the spatial variation of the growth rate $a(X)$ but is modified by the effects of diffusion. For instance, if $a(X)$ has the value $a$ in a segment of space, and the value $-\infty$ outside it, as shown first by Skellam \cite{Skellam}, $u(X)$ is given by the square of the elliptic $cd$ function \cite{KK, Skellam} . Thus $u(X)$, which is maximum at the center of the favorable region and drops off gradually towards the edges, is not proportional to $a(X)$. The latter has the shape of a step function. 

Equation (\ref{fisher}), which has been quite successful in the description of ecological dynamics \cite{Murray,Kot}, predicts interesting behavior in addition to the mere difference in shapes of $a(X)$ and $u(X)$. The maximum value $u_m$ of the density within the favorable region, i.e., the spatial segment in which $a(X)$ is positive, naturally depends on the length $L$ of the segment, and decreases if $L$ is decreased. What is striking is that there is a minimum value of $L$ below which $u_m$, and therefore the density throughout,  vanishes identically. This minimum  (critical) value can be obtained, except for a proportional constant equal to $\pi^2$, by equating the growth time $1/a$ to the diffusion time $L^2/D$:
\begin{equation}
L=L_s=\pi\sqrt{\frac{D}{a}}.
\label{dirichlet}
\end{equation}
The underlying physical idea is that the critical length of the favorable segment corresponds to the situation that the random walker (bacterium or animal) traverses the length of the segment diffusively in the time necessary for growth, and falls prey to the harsh conditions outside the segment.

Our interest in the present paper is to ask how this well-known result \cite{Skellam,Kot,KK} is modified by the simultaneous introduction of two items of realism into the growth $a(X)$:  a gradual, rather than step-like, spatial variation of the growth rate at the border of the segment, and a non-infinite value for destruction outside the segment. The latter, by itself, has been analyzed earlier  \cite{Ludwig} but the former has not, to the best of our knowledge. We provide an analysis of the combined situation below on the basis of wave function calculations from the time-independent Schr\"{o}dinger equation describing a quantum particle in a uniform gravitational or electrostatic field.

The analysis we present is applicable to at least two specific biological systems that have been studied recently: a bacterial population in a Petri dish, and rodents responsible for Hantavirus infection spread. In several experiments \cite{Lin,Perry} on the former system, the bacterial population is bathed in the presence of deadly ultraviolet
radiation, except in a localized region covered with a mask that protects the bacteria. The growth rate $a(X)$ is positive in the protected region and negative outside. In observations of the second system, there are patches of resources which make possible population growth and sustainance of rodents such as mice in parts of the landscape but not beyond those patches. Earlier analyses have treated the patches or the masks to correspond to \emph{sharply} changing resources. The real experimental situation is naturally different. The ultraviolet light that falls on the mask seeps under the edges of the
mask. As a consequence, the change in the growth rate is gradual rather than sudden. Similarly, the available water and food in the rodent patches  change gradually in the landscape. Transitions have been reported in experiments on the bacterial system \cite{Perry} following quantitative predictions made earlier \cite{KK}. Transitions also seem to be present in rodent experiments although the observations are more difficult to unravel. The observational relevance of our present investigation should be, thus, clear in both cases.

The paper is set out as follows. A simple model that incorporates a gradual variation of resources in space is presented in section 2 and analytical formulae are derived in terms of Airy functions. Limits are studied in section 3 and concluding remarks appear in section 4.

 \section{Model and the Result}

The model we analyze is as given in Eq. (\ref{fisher}) with the growth rate $a(X)$ as in Fig.1. It equals the constant $a$ for a spatial segment of length $L$ around the origin $X=0$, and drops gradually (linearly) to the constant negative value $-a_1$ outside the segment, the spatial variation being linear (constant slope). During the linear drop, the rate remains positive on either side of the segment for an additional distance $R$ as shown. Using the suffixes $1$, $2$ and $3$ for the respective regions I, II and III in Fig.1, we have
\begin{eqnarray}{\label{ne1}}
 \frac{\partial u_1(X,t)}{\partial t}&=&D\frac{\partial^2 u_1(X,t)}{\partial X^2}+au_1(X,t)-bu_1^2(X,t),\nonumber \\
\frac{\partial u_2(X,t)}{\partial t}&=& D\frac{\partial^2 u_2(X,t)}{\partial X^2}+a\left(1-\frac{2X-L}{2R}\right)u_2(X,t)\nonumber\\ &&-bu_2^2(X,t),\nonumber\\
\frac{\partial u_3(X,t)}{\partial t}&=&D\frac{\partial^2 u_3(X,t)}{\partial X^2}-a_1u_3(X,t)-bu_3^2(X,t).
\label{pde}
\end{eqnarray}

We do not attempt to solve either this nonlinear partial differential equation  or the corresponding nonlinear ordinary differential equation for the steady state $u(X)$ obtained by setting the time derivative equal to zero. Rather, we argue that, because our interest lies only in the extinction phenomenon, i.e., the vanishing of $u(X)$, we can eliminate the bilinear terms of the density, and match solutions of the ensuing \emph{linear} equation across the interfaces of the three regions I, II and III (see Fig. 1). This argument appears to have been given first by Ludwig et al. \cite{Ludwig} in his investigations of the  spatial patterning of the budworm. It has been used recently by the present authors to propose double-mask experiments for the bacterial problem \cite{KN}. The crux of the argument is that, at the transition, the densities vanish and therefore terms of order higher than the first may be safely neglected even while the linear terms are compared to extract the extinction condition. The critical value of $L$ signifying extinction which result for \emph{any positive} $b$ can therefore be calculated precisely by considering the linear equation obtained by putting $b$ equal to zero. The problem we are presented with here is simply of finding of the wavefunction for a time-independent Schr\"{o}dinger equation of a quantum particle in a potential which is linear in $X$ as in the case of a uniform gravitational or electrostatic field.

Defining the dimensionless position $x$ and the quantities $l$  and $r$, all obtained by dividing $X$, $L/2$, and $R$ respectively by the diffusion length $\sqrt{D/a}$, we focus on the linear steady state counterpart of Eq. (\ref{pde}), with the notation $\xi=\sqrt{a_1/a}$:
\begin{eqnarray}{\label{ne3}}
 &&\frac{d^2u_1(x)}{dx^2}+u_1(x)=0,\nonumber\\
 &&\frac{d^2u_2(x)}{dx^2}+\left(1-\frac{x-l}{r}\right)u_2(x)=0,\nonumber\\
 &&\frac{d^2u_3(x)}{dx^2}-\xi^2u_3(x)=0.
\end{eqnarray}
The growth rate plays the role of the potential in a 1-dimensional quantum mechanical problem in a constant force field. The first of these equations is solved trivially in terms of trigonometric functions, the third in terms of exponentials, and the second in terms of Airy functions.
\begin{figure}
 \includegraphics[width=8cm]{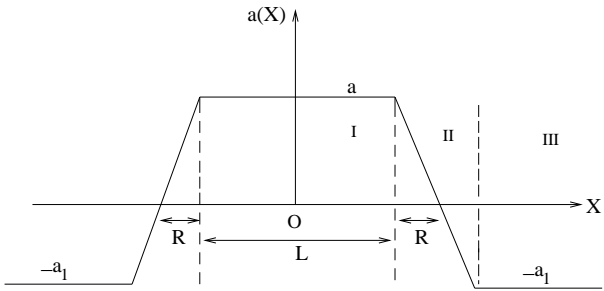}
 \caption{The spatial dependence of the growth rate $a(X)$ showing its value as the constant $a$ in a segment of length $L$ joined through a linear decrease to the value $-a_1$ on both sides. The width of the transition region is $R$ on either side.}
\end{figure}

Specifically, using the fact that the system has symmetry about $x=0$, the solution in the three regions is given by
 \begin{eqnarray}
  u_1(x)&=&A_1cos(x), \nonumber\\
 u_2(x)&=&A_2Ai\left[-r^{2/3}\left(1-\frac{x-l}{r}\right)\right]\nonumber\\ &&+B_2Bi\left[-r^{2/3}\left(1-\frac{x-l}{r}\right)\right], \nonumber\\
u_3(x)&=&A_3e^{-\xi x},
 \label{solns}
  \end{eqnarray}
where $Ai$ and $Bi$ are Airy functions and the $A$'s and $B_2$  are constants to be determined from the boundary conditions.

Matching the logarithmic
derivative of the solutions across the interfaces and introducing $\eta=r^{1/3}$ for notational convenience, we arrive at the following condition for extinction.
\begin{widetext}
 \begin{equation}
 \tan l=\left(\frac{1}{\eta}\right)\left(\frac{Ai^{\prime}(\xi^2\eta^2)Bi^{\prime}(-\eta^2)-Ai^{\prime}(-\eta^2)Bi^{\prime}(\xi^2\eta^2)+\xi\eta
\left[Ai(\xi^2\eta^2)Bi^{\prime}(-\eta^2)-Bi(\xi^2\eta^2)Ai^{\prime}(-\eta^2)\right]}
   { Ai(-\eta^2)Bi^{\prime}(\xi^2\eta^2)
-Ai^{\prime}(\xi^2\eta^2)Bi(-\eta^2)+\xi\eta\left[Ai(-\eta^2)Bi(\xi^2\eta^2)-Ai(\xi^2\eta^2)Bi(-\eta^2)\right]}\right).
\label{central}
 \end{equation}
\end{widetext}
Primes denote derivatives. The normalized extinction length $l=L/\sqrt{D/a}$ is the arctangent of the right hand side if the latter is positive, and is zero otherwise. 

Equation (\ref{central}) is the central result of this paper. The crucial parameters in the expression are $\eta$ which measures the transition region relative to the diffusion length and $\xi$ which measures the death rate outside, relative to the growth rate inside, the segment. As is clear from the expression, the second parameter can also be chosen naturally as the product $\xi \eta$ . 

\section{Limiting cases}
If the transition region vanishes ($r=0$), we take the limit $\eta \rightarrow 0$ in Eq. (\ref{central}) and recover the Ludwig formula \cite{Ludwig}
 \begin{equation}
L=L_d=2\sqrt{\frac{D}{a}} \arctan{\sqrt{\frac{a_1}{a}}}.
\label{redludwig}
\end{equation}
In the general case $r \neq 0$, the extinction value normalized to this Ludwig value is given by plotting the analytic expression in Eq. (\ref{central}). Fig. 2 shows, in the form of the solid line in the main figure, this dependence of $L/L_d$ on $r$, the normalized extent of the transition region, for a specific value of $\xi$, viz., 3. The solid circles are obtained via the numerical solution of the full  nonlinear equation (\ref{pde}), for a nonzero $b=1$ in appropriate units. Note that Eq. (\ref{pde}) cannot be solved analytically, and that the complete agreement between the numerical results from the nonlinear equation (whatever the value of $b$) and the analytic result from the linear extinction analysis (that has no $b$) vindicates our theoretical procedure. 

\begin{figure}
 \includegraphics[bb=50 50 224 172]{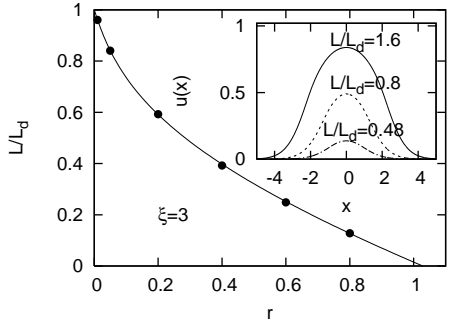}
 \caption{ Excellent agreement between our Airy extinction prediction formula (\ref{central}) based on the linear theory (solid line) and the numerical solution of the full nonlinear equation (circles). Plotted in the main figure is the extinction length normalized to the Ludwig value, $L/L_d$, versus the extent of the transition region $r$ for $\xi=3.0$. The inset shows steady state density profiles corresponding
 to three different values of $L$ larger than the critical, for $\xi=3.0$ and $r=0.4$. Distance $x$ is expressed in units of 
 diffusion length $\sqrt{D/a}$ while the steady state density $u(x)$ is expressed in units of $a/b$.  The density decreases as the patch length decreases and  vanishes at the critical value, at which $L/L_d=0.398$.}
\end{figure}

Various limiting cases may be studied from our Eq. (\ref{central}) through ascending and asymptotic expansions of the Airy functions. An example of the former is \cite{Abrom,Vallee}
  \begin{eqnarray}
   Ai(x)&=&c_1f(x)-c_2g(x),\nonumber\\
   \frac{Bi(x)}{\sqrt{3}}&=&c_1f(x)+c_2g(x),
  \end{eqnarray}
  where,
  \begin{eqnarray}
   f(x)&=&1+\frac{1}{3!}x^3+\frac{1.4}{6!}x^6+...\nonumber\\
   g(x)&=&x+\frac{2}{4!}x^4+\frac{2.5}{7!}x^7+....\nonumber
  \end{eqnarray}
  and $c_1=Ai(0)$, $c_2=Ai^{\prime}(0)$.
  
  One of the goals of the present paper is the generalization of the Skellam result (\ref{dirichlet}) in the presence of a transition region. For this, we let $\xi \rightarrow \infty$ in (\ref{central}) and obtain
 \begin{eqnarray}{\label{skelgen}}
  L=2\sqrt{\frac{D}{a}} \arctan \left[-\frac{1}{r^{1/3}}\frac{Ai^{\prime}(-r^{2/3})}{Ai(-r^{2/3})}\right].
 \end{eqnarray}

\begin{figure}
 \includegraphics{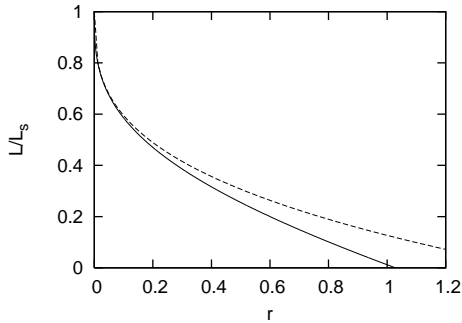}
 \caption{Generalization of the Skellam result (\ref{dirichlet}) in the presence of a transition region of extent $R$. We plot $L/L_s$ as a function of $r=R/\sqrt{D/a}$. The solid line is the exact result (Eq. \ref{skelgen}) and the dashed line the approximation retaining the lowest power of $r^{1/3}$. Note that $L=0$ for $r>1.028$.}
\end{figure}

 We divide Eq. (\ref{skelgen}) by the Skellam result (\ref{dirichlet}), and plot $L/L_s$ as a function of the transition region extent (i.e., $r=R/\sqrt{D/a}$) in Fig. 3. The exact expression is the solid line. It is approximated reasonably well for small $r$ by the ascending series approximation
 $$L=\sqrt{\frac{D}{a}}\left(\pi-2c r^{1/3}\right)$$
denoted by a dashed line, with $c=-Ai(0)/Ai^{\prime}(0)=1.372$. Important to note is that the exact $L$ becomes zero at a finite value of the transition region extent (at $r=1.028$). This follows from the existence of a zero of $Ai^{\prime}(-r^{2/3})$ at finite $r$; the physical significance is commented on below.
 
\begin{figure}[h]
 \includegraphics[width=9.7cm,bb=50 50 403 172]{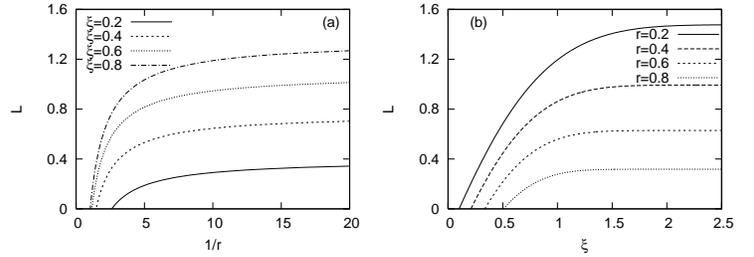}
\caption{Our analytic predictions for the dependence of the extinction length $L$ (in units of $\sqrt{D/a}$) on the transition region extent in (a) and against the ration of the death to the growth rate in (b), plotted from the Airy function formula Eq. (\ref{central}). Note the threshold values of $1/r$ and $\xi$ below which extinction does \emph{not} occur.}
\end{figure}

 The extinction length $L$ given by the exact expression (\ref{central}) is plotted in Fig. 4 against (a) the reciprocal of the transition region extent, $1/r$, for different values of $\xi$ as shown, and (b) the square root of the ratio of the death rate to the growth rate, $\xi$ for different values of $r$. The extinction length becomes zero for large enough $r$ in (a) and small enough $\xi$ in (b). This means that there is no extinction for $r>r_c$ and $\xi<\xi_c$ where $r_c$ and $\xi_c$ are the threshold values. These threshold values are plotted in Fig. 5 directly from the graphs. We have verified that the curves coincide precisely with those obtained by putting the numerator of Eq. (\ref{central}) equal to zero.

\begin{figure}[h]
 \includegraphics[width=9.7cm,bb=50 50 403 172]{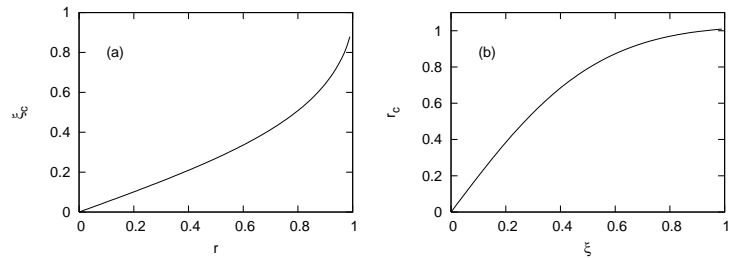}
\caption{Plots of threshold values $\xi_c$ and $r_c$ obtained graphically. They are found to agree with
 the prediction of Eq. (\ref{central}) obtained by putting its numerator equal to zero.}
\end{figure}

 This vanishing of the extinction length $L$, signifying that the extinction phenomenon disappears for large enough transition region  or small enough death rate, stems from the definition of $L$ as the top segment of the $a(x)$ trapezoid in Fig. 1. Thus, Fig. 6 shows with this segment of zero extent  ($L=0$)  that the favorable patch is still wide enough to make possible for the walkers to survive in the protected patch because of the transition region.

\begin{figure}[h]
 \includegraphics[width=8cm]{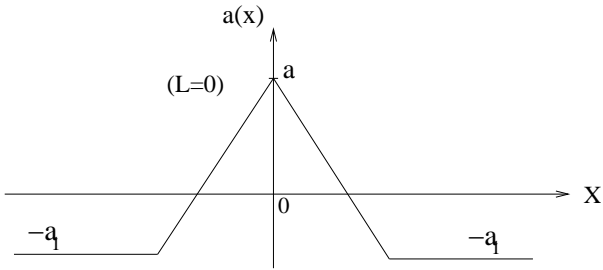}
\caption{Situation where $L=0$, there being no extinction because the transition region itself provides a large enough favorable patch for survival.}
\end{figure}

\section{Conclusion}
Our focus in this paper has been to study the combined effect of a gradual change in resources at the borders of a patch, and of finite death rates outside the patch, on the extinction transition of bacteria in a Petri dish, rodents in a landscape, or of similar random walkers that can move over an inhomogeneous spatial region. Although the system is essentially nonlinear, and the extinction transition is a consequence of the nonlinearity (coupled to the inhomogeneity), we have been successful in obtaining exact solutions through an analytic procedure. The procedure draws on an analogy to the solution of a time-independent Schr\"{o}dinger equation and results in an expression for the critical length of the patch in terms of Airy functions and their derivatives. 

Our central result is Eq. (\ref{central}). One of the crucial parameters is $\eta$ which is $r^{1/3}$, where $r$ is the length of the transition region in units of the diffusion length $\sqrt{D/a}$. The appearance of the latter is natural since the underlying process is the movement of the random walker (the diffusing entity) from within the favorable patch to the unfavorable space outside within the growth time. The emergence of the third power is a consequence of the Airy function which arises from the linear dependence of the gradual decrease of the resources.

The other quantity of importance is the ratio of the death rate outside the patch to the growth rate inside, specifically $\xi=\sqrt{a_1/a}$. This second parameter can also be taken to be the combination $\xi \eta=\sqrt{r^{2/3}a_1/a}$. We have recovered the Skellam result  Eq. (\ref{dirichlet}) and given its generalization in Eq. (\ref{skelgen}).  We have also recovered the Ludwig result and provided its generalization as well.

This work was supported in part by the NSF under
grant no. INT-0336343, and by NSF/NIH Ecology of Infectious Diseases under grant no. EF-0326757.

\end{document}